\documentstyle[emulapj,epsfig]{article}

\begin{document}

\bigskip

\font\two=cmbx10 scaled \magstep1

\title{\bf Oscillations of MHD shock waves on
the surfaces of  T~Tauri stars}

\author{A.V. Koldoba}
\affil{Institute of Mathematical Modelling, Russian Academy of
Sciences, Moscow, Russia; ~ koldoba@rambler.ru}

\author{G.V. Ustyugova} \affil{Keldysh Institute of Applied Mathematics, Russian Academy of
Sciences, Moscow, Russia; ~ ustyugg@rambler.ru}

\author{M. M.~Romanova}
\affil{Department of Astronomy, Cornell University, Ithaca, NY
14853-6801; ~ romanova@astro.cornell.edu}

\author{R. V.E.~Lovelace}
\affil{Department of Astronomy and Applied and Engineering Physics,
Cornell University, Ithaca, NY 14853-6801; ~RVL1@cornell.edu }

\def\dst{\displaystyle}        
\def\tst{\textstyle}           %
\def\sst{\scriptstyle}         %
\def\ssst{\scriptscriptstyle}  %

\def\const{{\rm const}}        
\def\cos{\mathop{\rm cos}\nolimits}
\def\sin{\mathop{\rm sin}\nolimits}
\def\tg{\mathop{\rm tg}\nolimits}
\def\ctg{\mathop{\rm ctg}\nolimits}
\def\matharrow{\mathop{\protect{\relbar\joinrel\longrightarrow}}\limits}

\def\arh#1{\renewcommand{\arraystretch}{#1}}
\def\arhp{\renewcommand{\arraystretch}{1.5}}

\def\prodi#1#2{\frac{d #1}{d #2}}
\def\prodii#1#2{\frac{d^2 #1}{d {#2}^2}}

\def\parti#1#2{\frac{\partial #1}{\partial #2}}
\def\partii#1#2{\frac{\partial^2 #1}{\partial {#2}^2}}

\begin{abstract}

   This work treats the matter deceleration in a
magnetohydrodynamics
radiative shock wave at the surface of a star.
  The problem is relevant
to classical T~Tauri stars where
infalling matter is channeled along the star's magnetic
field  and  stopped in the dense layers of photosphere.
   A significant new aspect of the present work is that the magnetic
field  has an arbitrary angle with
respect to the normal to the star's surface.
    We consider the limit where
the magnetic field at the surface of the
star is not very strong in the sense that
the inflow is super Alfv\'enic.
   In this limit the initial
deceleration and heating of plasma (at the entrance to the cooling
zone) occurs in a fast magnetohydrodynamic shock wave.
   To calculate the intensity
of radiative losses we use ``real" and  ``power-law"
radiative functions.
    We determine the stability/instability of the
radiative shock wave as a function of parameters of the incoming
flow: velocity, strength of the magnetic field, and its inclination
to the surface of the star.
     In a number of simulation runs with the
``real" radiative function, we find a simple criterion for
stability of the radiative shock wave.
    For a wide range
of parameters, the periods of oscillation of the shock wave are of
the order $0.02-0.2 ~{\rm s}$.

\end{abstract}


\section{Overview of the Problem}

In classical T Tauri stars (CTTSs) matter accretes from the disk to
a star through magnetospheric funnel streams (Camenzind 1990;
K\"onigl 1991; see also recent review by Bouvier et al. 2007).
Similar type accretion but at smaller scales is expected to a
magnetized white dwarf (e.g., Warner 1995) and magnetized neutron
stars (e.g., Ghosh \& Lamb 1979). In the funnel stream, matter is
lifted above the equatorial plane and falls down onto the star due
to the gravitational acceleration.   Large-scale magnetospheric flow
has been recently investigated in 2D magnetohydrodynamic (MHD)
simulations (Romanova et al. 2002; Bessolaz et al. 2008) and in full
3D MHD simulations (Romanova et al. 2003, 2004; Kulkarni \& Romanova
2005).  Many aspects of the global magnetospheric flow are now
understood. However, interaction of the funnel streams with the
surface of the star has not been adequately investigated.


\begin{figure*}[t]
\epsscale{1.0} \plotone{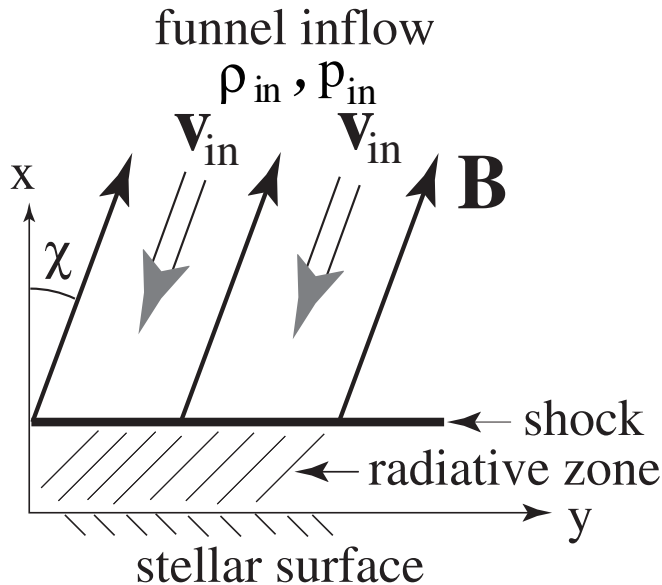} \caption{Sketch of the geometry of
the flow.
  Matter with density
$\rho_{\rm in}$ and pressure $p_{\rm in}$ flows towards the star
with velocity $v_{\rm in}$ along the magnetic field which has
strength $B_{\rm in}$ and is inclined relative to the normal to the
surface of the star $x$ at an angle $\chi$. Matter slows down in the
MHD shock wave close to the surface of the star and radiates and
cools down in the ``cooling zone". Matter may have different angle
relative to the field, but we consider the coordinate system in
which vectors velocity and magnetic field are parallel to each
other.} \label{Figure 1}
\end{figure*}


Theoretical models indicate that  close to a star matter in the
funnel streams is accelerated  to almost free-fall velocity, before
it hits the high-density layers of the stellar atmosphere. The
matter rapidly slows down,  forming a shock wave close to the
stellar surface.
   Most of the energy of the flow is radiated behind the shock
wave  (e.g., Lamzin 1995, 1998; Muzerolle et al. 1998; Calvet \&
Gullbring 1998; Gullbring et al. 2000; Ardila \& Basri 2000).
      In the case of CTTSs most of energy is radiated in the ultraviolet and soft
X-ray bands (e.g., Calvet \& Gullbring 1998; G\"unther \& Schmitt
2008).   The 3D MHD simulations of  magnetospheric flow have show
that the hot spots are inhomogeneous and are expected to have higher
temperature in the central regions of spots compared to peripheral
regions (Romanova et al. 2004). This may have a number of important
consequences for investigation of hot spots including the dependence
of the filling factor on the wavelength.

If  star's magnetic field of the star is {\it strong},  then close
to the stellar surface the magnetic energy-density is larger than
that the kinetic energy-density of the matter.
    Consequently the matter is passively channeled along the field
lines. In this sub-Alfv\'enic regime a hydrodynamic approach is
usually adopted for modeling the shock waves (see \S 2).
    The shock wave is found to be non-stationary.   It {\it oscillates} with
a high frequency  due to the competition
between accretion heating in the shock front and radiative cooling
behind the front (Langer, Chanmugan \& Shaviv 1981; Chevalier \&
Imamura 1982).
       If the magnetic field is not very strong near the
surface of the star then the flow may be super-Alfv\'enic.
 In this regime the orientation of the magnetic
field may influence stability of the shock.
        Only the special case of the flow perpendicular to the
magnetic field has been considered so far.
    It is know that this transverse magnetic field can suppress
instability of the shock wave (Smith 1989; Toth \& Draine 1993).
          In this paper we investigate the stability of the radiative shock waves in
the super-Alfv\'enic regime for different orientations of the magnetic
field relative to stellar surface.
           We consider small patch  of the hot
spot and investigate the stability of the radiative MHD shock wave
for parameters typical for CTTSs.
            Figure 1 shows a sketch of the considered geometry.

For CTTSs the expected periods  of the oscillations of the shock are
very short.   The periods vary between $0.02$ and $0.2$ seconds
depending on the parameters. Oscillations in this period range have
not been observed so far.  Smith, Jones \& Clarke (1996) searched
for rapid photometric variability in several CTTSs in the range of
periods from minutes to hours and did not find oscillations.    Much
higher temporal resolution is required to resolve the oscillations
discussed in this paper.

Section 2 of this paper discusses the earlier research on radiative
shocks.
    Section 3 discusses the model and basic equations,
and Section 4 the dimensionless variables and scalings.
  Section 5 comments on the dimensionless variables and the
scalings of different quantities, and also describes the stationary
structure of the shock. Section 6 discusses the methods used to
study the time-dependent shocks, and Section 7 gives our results.
Section 8 gives the conclusions of this work.

\section{Earlier Research of Radiative Shocks and Radiative Cooling Function}

 The stability of shock waves has been investigated by different groups
both analytically (linear analysis) and numerically. Most of the
investigations have been restricted to  a {\it purely hydrodynamic
analysis}, because in many situations matter is channeled along the
field lines and the problem can be considered as non-magnetic.

First results on interaction of the funnel streams with a star and
cooling in the radiative shock wave have been obtained in
application to accreting white dwarfs. Numerical modeling of
radiative shock wave at the surface of white dwarf led to discovery
of instability which is driven by alternation of accreting heating
in the shock wave and radiative cooling behind the shock wave and in
resulting oscillations of the position of the shock front
 (Langer, Chanmugan \& Shaviv 1981). Linear analysis of
the stability of a one-dimensional radiative shock wave was done by
Chevalier \& Imamura (1982).
   These authors assumed that the
radiative cooling from a unit volume is $\rho^2 T^{\alpha}$, and
they studied the stability  as a function $\alpha$.
   In a more general form the problem has been investigated by
Ramachandran \& Smith (2004).
   This work assumed that the
radiative losses vary as $\rho^{\beta} T^{\alpha}$.
  The  boundaries of the stable and unstable regions were found
as well as the frequencies and growth rates of the lowest frequency
modes for different values of $\alpha,~ \beta$.
   Ramachandran \& Smith (2006)
investigated the influence of the Mach number of the inflowing gas
to the stability of the radiative shock wave. In addition they
considered flow at different adiabatic indices $\gamma$ typical for
astrophysical applications and two types of the boundary conditions
at the ``wall" where the flow is stopped.
   The influence of boundary conditions on the stability
of  radiative shock waves has been investigated in detail by Saxton
(2002).

    A detailed investigation of the linear and nonlinear evolution of
one-dimensional radiative shock waves has been done by Mignone
(2005, see also Toth \& Draine 1993). The intensity of the radiative
losses was taken to be $\rho^2 T^{\alpha}$.
    The stability has been investigated in the
linear approximation for the first eight low-frequency modes and it
has been established, that:
 (1) The first eight modes are
stable for $\alpha > 0.92$; (2) The fundamental mode ($n=0$)
becomes unstable for  $\alpha < 0.388$,
the $n=1$  mode for $\alpha < 0.782$;
(3) In the unstable regime the growth rate is larger for larger mode
numbers, but the growth rate of the higher-$n$ modes is not
very different from the growth rate of the $n=7$ mode;
(4) The normalized frequencies of the corresponding modes have
an approximately linear dependence on $n$, that is, $\omega_n(\alpha)
= \omega_0(\alpha) + n \Delta \omega(\alpha)$, and decrease
as $\alpha$ increases (excluding the fundamental mode $n=0$).

\begin{figure*}[t]
\epsscale{0.9} \plotone{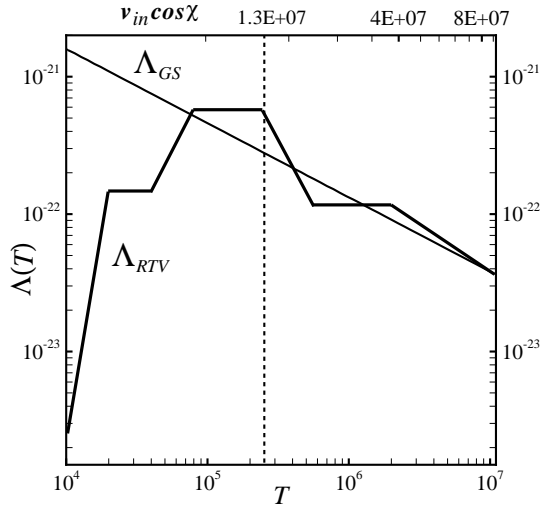} \caption{Cooling function
$\Lambda_{\rm RTV}(T)$ (thick solid line) and $\Lambda_{\rm GS}(T)$
(thin solid line) as a function of $T$ in K. The vertical dotted
line shows our illustrative value of velocity before the shock
$v_{in}\cos\chi = 1.3\times 10^7 {\rm cm~s^{-1}}$ and the
corresponding temperature behind the shock.} \label{Figure 2}
\end{figure*}


The stability of a radiative shock wave in the presence of a {\it
magnetic field} was investigated by Smith (1989).
   A more  detailed investigation of the shock stability
in presence of a magnetic field was done by Toth \& Draine (1993;
hereafter TD93).
  For the situation considered, the matter flows onto the shock
wave perpendicular to its front and the magnetic field has only
component parallel to the front.
   It was concluded that even a
modest magnetic field may lead to stabilization of a radiative
shock wave which is unstable in the hydrodynamic limit.
   The higher the harmonic number, the larger the value of the magnetic field
which is needed for stabilization of the front for a fixed value of
$\alpha$.
   First of all, the magnetic field stabilizes the
fundamental mode.
  In particular, for $\alpha = -0.5$ the fundamental
mode is stabilized at $ M_A^{-1} = [( B_y/\sqrt{4 \pi\rho})/v]_{in}
 = 0.15$, while the higher modes are stabilized at $M_A^{-1}
= 0.5$, where $v$ is velocity of the incoming flow and $B_y$  is
magnetic field parallel to the front.

The subsequent investigation of the stability of the radiative
shocks with transversal magnetic field has been done by
Ramachandran \& Smith, (2005; hereafter RS05). They considered
different values of $\alpha,~ \beta$ in the dependence
of the radiative
losses, different values $\gamma$ for the adiabatic index, and also
different Mach numbers in the inflowing matter.
   They obtained new
results, and also rederived accurately results by TD93 for the
case where $\gamma = 5/3$ and $\beta = 2$.

   These earlier studies show that the stability of  radiative
shock waves  depends strongly on the
functional dependence of the radiative cooling.
   In the present work we consider
$\beta = 2$  and a monatomic
gas with $\gamma = 5/3$.
   We  investigate the stability of
the radiative shock waves using a ``real"
radiative loss function.
   We calculate this ``real"
function and approximate it with the power laws.
    We assume that there
is collisional ionization equilibrium (CIE).
   In this approximation  photons
freely escape the plasma.
   Thus, full
thermodynamic equilibrium is not established, and the Saha's formula
(for calculation of the degree of the ionization) is not applicable
(e.g., Spitzer 1968). The radiative losses under these conditions
have been calculated by a number of authors.
  In these
calculations the abundances of the elements are assumed to be
Solar.
   In the paper by Rosner, Tucker, \& Vaiana (1978; hereafter
RTV) based on the calculations of Raymond \& Smith (1977), the
radiative losses in the temperature interval $10^{4.3}{\rm K} < T <
10^7 {\rm K}$ are $n_e n_H \Lambda(T)$, where $n_e, n_H$ are
electron density and hydrogen density (total), while the radiation
function $\Lambda(T)$ has been approximated by a multi-segmented
power-law (see below). Subsequently, $\Lambda(T)$ was been
calculated at the interval $10^{3.65}{\rm K} < T < 10^8 {\rm K}$
(Peres et al, 1982).
  We refer to this function as the ``real"
cooling function and add a subscript $RTV$.
   The dependence is
 $$
\Lambda_{\rm RTV}(T) = \left\{
\begin{array}{lc}
(10^{-7.85} T)^{6.15} & 10^{3.9} {\rm K} < T < 10^{4.3} {\rm K}\\
10^{-21.85}           & 10^{4.3} {\rm K} < T < 10^{4.6} {\rm K}\\
10^{-31} T^2          & 10^{4.6} {\rm K} < T < 10^{4.9} {\rm K}\\
10^{-21.2}            & 10^{4.9} {\rm K} < T < 10^{5.4} {\rm K}\\
10^{-10.4} T^{-2}     & 10^{5.4} {\rm K} < T < 10^{5.75} {\rm K}\\
10^{-21.94}           & 10^{5.75} {\rm K} < T < 10^{6.3} {\rm K}\\
10^{-17.73} T^{-2/3}  & 10^{6.3} {\rm K} < T < 10^{7} {\rm K}\\
10^{-18.21} T^{-0.6}  & 10^{7} {\rm K} < T < 10^{7.6} {\rm K}
\end{array}
\right.
$$

   The most recent results for radiative
losses in the CIE-approximation are
given by Gnat \& Sternberg (2007; hereafter GS07).
   In this work it
was accepted that the relative abundances of the hydrogen and helium
are $n_{\rm He}/n_{\rm H} =1/12$. Abundances of other elements
(${\rm C, Ni, O, N, Mg, Si, S, Fe}$) are small and they do not give
a contribution to the total pressure. However, the intensity of the
radiative losses significantly depends on the relative abundance of
these elements. For the Solar abundance GS07 proposed an
approximation formula:
 \begin{equation}
\Lambda_{\rm GS} = 2.3 \times 10^{-19} T^{-0.54}~~~~{\rm erg~
cm^3~s^{-1}}~, \label{1}
\end{equation}
in the temperature range $10^5K < T < 10^8 {\rm K}$.
    The left
boundary of this interval corresponds approximately to the maximum
of the ``real" radiation function which is  at $2.3 \times
10^5 {\rm K}$.
  For temperatures higher than $6 \times 10^7 {\rm K}$,
the dominant mechanism of radiative losses is bremsstrahlung
radiation with the temperature dependence $\Lambda \sim \sqrt{T}$.

According to calculations of GS07 in
CIE-approximation,  hydrogen is completely ionized for
temperature $T > 3 \times 10^4 {\rm K}$, and Helium for
$T > 2 \times 10^5 {\rm K}$.
   Accepting their abundances discussed above we
obtain for $T > 2 \times 10^5 {\rm K}$, the electron density $n_e =
n_{\rm H} + 2 n_{\rm He}$, and an average mass per particle is
$0.6m_p$.
   Thus, the model which we use (based on the
CIE-approximation) is applicable for $T > 2 \times 10^5 {\rm K}$.
At lower temperatures, the partial ionization
of Helium and the change of the
average mass per particle become significant.
   At the temperature $T
= 3 \times 10^4 {\rm K}$ the fractions of the neutral and ionized
atoms of  hydrogen and helium are: $x({\rm H})=3.6 \times
10^{-3},~x({\rm H^+})=0.997,~x({\rm He})=0.4,~x({\rm
He^+})=0.6,~x({\rm He^{++}})=0$. Thus an average mass per particle
is $0.62 m_p$.
   At a temperature $T =
2 \times 10^4 {\rm K}$, the corresponding fractions are $x({\rm
H})=0.078,~x({\rm H^+})=0.922,~ x({\rm He})=0.993,~x({\rm
He^+})=0.007,~x({\rm He^{++}})=0$, where an average mass per
particle is $0.66 m_p$. We neglect this factor.

Subsequently we assume that plasma is an ideal gas with equation of
state $p={\cal R} \rho T$ where ${\cal R} = k_B/(0.6 m_p) = 1.385
\times 10^8 {\rm erg~g^{-1}~K^{-1}}$ is gas constant.

In the paper GS07 it is shown that for $T < 10^6 {\rm
K}$, the radiative losses, calculated in CIE-approximation, exceed
losses which occur in  non-stationary plasma cooling.
   This is connected with the fact that
in the first case the ionization level
of elements participating in the main radiative processes is lower
because in the non-stationary regime recombination lags the
cooling. However, we  use the more detailed approximation mentioned
above after changing the normalization.

   Figure 2 shows the radiative cooling functions $\Lambda_{\rm RTV}(T)$
(thick solid line) and $\Lambda_{\rm GS}(T)$ (thin solid line).
   The diagonal dashed line shows the dependence of the
upstream  velocity normal to the shock ($v_{in}\cos\chi$)
on  the temperature behind the shock (for $\gamma=5/3,
p_{\rm in} = 0$):
$ T_s = 3 (v_{\rm in} \cos \chi)^2/(16
{\cal R})$.
   The horizontal dashed
line shows the velocity of the incoming flow $v_{\rm in}$ (for
illustration of the calculated results we take $v_{\rm in} \cos \chi
= 1.3 \times 10^7 {\rm cm~s^{-1}}$) and corresponding to this
velocity temperature behind the shock wave front.

    Section 2 of the paper discusses the model and basic equations,
and Section 3 the dimensionless variables and scalings.
  Section 3 comments on the dimensionless variables and the
scalings of different quantities.  Section 4 describes the
stationary structure of the shock.   Section 5 discusses
the methods used to study the time-dependent shocks, and
Section 6 gives our results.  Section 7 gives the conclusions
of this work.

\section{Formulation of the Problem}

We investigate formation and evolution of the shock wave near the
surface of the star which forms as a result of disk accretion to a
star through a funnel flow. The accreting matter is sufficiently
ionized to satisfy the frozen-in condition so that matter of the
funnel streams is channeled by the magnetic field and close to the
star it flows along the field lines.

Figure 1 shows the geometry.
   Matter with velocity
$v_{\rm in}$, density $\rho_{\rm in}$, and pressure $p_{\rm in}$
flows towards the surface of the star along the magnetic field.
   The magnetic field has a strength $B_{\rm in}$ and is
directed at an angle $\chi$ relative to the normal vector to the
surface of the star $\hat{\bf x}$. The $x$-axis is normal to the
surface of the star and the $y-$axis is tangential to its surface
and is directed such that the magnetic field is located in the
$(x,y)$ plane. We neglect small perturbations in $z-$direction
associated with propagation of the Alfv\'en waves. We consider that
the  magnetic field has an  arbitrary inclination angle $\chi$
relative to the normal
 to the star's surface.
    In general, the matter  flow velocity is not
parallel to the magnetic field.
 However, such parallel orientation can be obtained by transformation of
 the coordinate system.

Heating of matter occurs in the front of the MHD shock wave.
   In the cooling zone behind the shock, matter radiates
energy, is decelerated, and become denser up to the moment, when the
radiative cooling stops.
   Formally this happens when $T=0$.
The height of the radiative zone is small compared to either width
of the funnel stream or to the radius of the star.
  Thus we can neglect the
inhomogeneity of the accretion flow in the $(y,z)$ directions. (See
Canalle et al. 2005 for a discussion of cases where the converging
of the field lines is important.)
    We also neglect small
effects associated with acceleration by gravity of the star
(considered e.g., by Cropper et al. 1999) because the  region we
consider is small.
  Generally, the structure consisting of the shock wave and
the  cooling zone is unstable  to both longitudinal, $x$)
perturbations (which can be studied in one-dimensional approach),
and to transverse ($y,z$) perturbations (Bertschinger 1986; Imamura
et al. 1996).
  The latter makes the problem more than
one-dimensional even if the flow is homogeneous.
   In the present work
the spatial perturbations are not considered.

  We consider conditions where
the star's mass is $M_* =
0.8M_{\sun}$ and its radius is $R_* = 2 R_{\sun}$.
   In this case the
free-fall speed at the star's surface is $ v_{ff} = \sqrt{2GM_*/R_*}
= 4 \times 10^7 {\rm cm~s^{-1}}$.
   If the temperature
of accreting matter is $10^4 {\rm K}$, then the sound speed is $2
\times 10^6 {\rm cm~s^{-1}}$ (for average mass per particle
$0.6m_p$).
   Clearly, the accretion is strongly
supersonic;   the sonic Mach number is $20$.
     For a surface magnetic field
$B=10^3 {\rm G}$ and density of the inflowing matter $10^{-11} {\rm
g~cm^{-3}}$ (Romanova et al. 2002),  the Alfv\'en velocity is $ c_A
= B/\sqrt{4 \pi \rho} = 10^8 {\rm cm~s^{-1}}$.
     For these
parameters the flow  is sub-Alfv\'enic.
   However, for $B < 3\times 10^2 {\rm G}$  and the other parameters
the same the flow is super-Alfv\'enic
and perturbations from the shock wave cannot propagate up the
stream.
   Strictly speaking we should compare the velocity of the
accreting matter with the fast magnetosonic velocity, but it does
not differ significantly from the Alfv\'en velocity because the sound speed
is small.
   Thus, both cases are interesting: stability of the
radiative shock waves for sub-Alfv\'enic
and super-Alfv\'enic inflows.
  However, we restrict the present study to
super-Alfv\'enic inflow.
\begin{figure*}[t]
\epsscale{1.1} \plotone{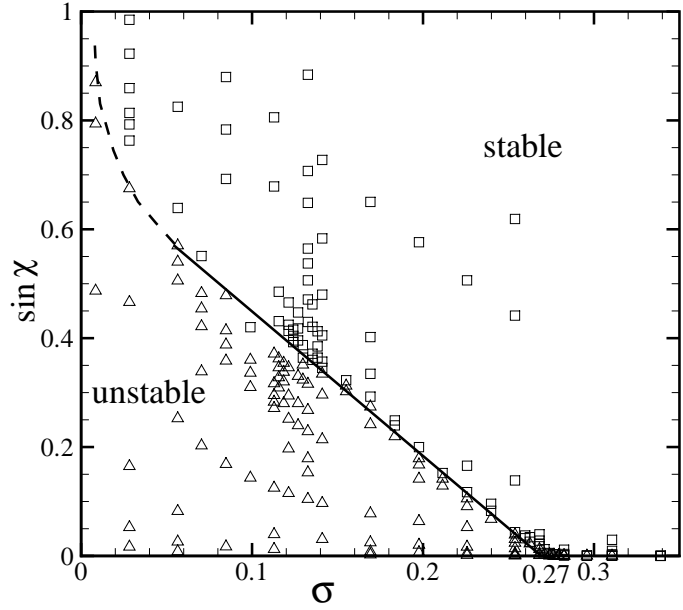} \caption{The position of the
boundary between stable and unstable radiative shock waves as a
function of parameters $(\sigma, \sin \chi)$ for a velocity of
accretion $v_{\rm in} \cos\chi = 1.3 \times 10^7 {\rm cm~s^{-1}}$.}
\label{Figure 3}
\end{figure*}

We assume that the radiation zone behind the shock
is optically thin in the
direction perpendicular to the star's surface and to the front
of the shock wave.
   The dominant radiation loss mechanisms
are determined by the ``thermodynamic" state of plasma.
    We assume
that the state of the matter can be
described in terms of a temperature
which is the same for the ions and electrons.
   For the considered conditions,
the radiative losses per unit volume is
$\rho^2 \Lambda(T)$, where $\rho$ is the plasma density and
$\Lambda(T)$ is the radiative function obtained from $\Lambda_{\rm
RTV}(T)$ or from $\Lambda_{\rm GS}(T)$ by renormalization.
   We consider
that the accreting matter is an ideal gas with adiabatic index
$\gamma = 5/3$ both before the compression at the shock front and
in the radiation zone.

In the absence of the magnetic field, and when the matter falls
perpendicular to the surface of the star, the temperature behind the
shock  is $T = 10^5 v_7^2 ~{\rm K}$, where $ v_7 = v/(10^7 {\rm
cm~s^{-1}})$, and the average mass per particle is $0.6m_p$.
   For velocities of the incoming matter  $v = 4
\times 10^7 {\rm cm~s^{-1}}$, the temperature behind the shock
reaches $1.6 \times 10^6 {\rm K}$.  At such temperature both
hydrogen and helium atoms are completely ionized according to the
CIE approximation.

   We consider the situation where the matter accretes to the star along
magnetic field lines inclined by an angle $\chi$ to the normal to
the star's surface.
  The  shock is parallel to the star's surface and it remains
parallel as it moves.
    That is, we consider variations only in the $x-$direction.
   The equations describing this situation are
following:
 \begin{eqnarray}
\parti{\rho}{t} + \parti{(\rho v_x)}{x} = 0~,
\nonumber
\end{eqnarray}
\begin{eqnarray}
\parti{(\rho v_x)}{t} + \parti{}{x} \left( \rho v_x^2 + p +
\frac{B_y^2}{8\pi} \right) = 0~,
\nonumber
\end{eqnarray}
\begin{eqnarray}
\parti{(\rho v_y)}{t} + \parti{}{x} \left( \rho v_x v_y -
\frac{B_x B_y}{4\pi} \right) = 0~,
\nonumber
\end{eqnarray}
 \begin{eqnarray}
  B_x  = \const~,
\nonumber
\end{eqnarray}
\begin{eqnarray}
\parti{B_y}{t} + \parti{}{x} \left( v_x B_y - v_y B_x \right) = 0~,
\nonumber
\end{eqnarray}
 \begin{eqnarray}
\parti{}{t} \left( \frac{\rho v^2}{2} + \frac{p}{\gamma-1} +
\frac{B^2}{8\pi} \right)
\nonumber
\end{eqnarray}
\begin{eqnarray}
\quad+ \parti{}{x} \left[ v_x \left( \frac{\rho v^2}{2}+ \frac{\gamma
p}{\gamma-1} \right) + \frac{B_y}{4\pi} \big( v_x B_y - v_y B_x
\big) \right]
\nonumber
\end{eqnarray}
\begin{equation}
\quad \quad \quad= - \rho^2 \Lambda(T)~.
\end{equation}
  Here $v^2 \equiv v_x^2 + v_y^2~,~B^2 \equiv B_x^2 + B_y^2$.
In the unperturbed flow upstream of the shock,
\begin{eqnarray}
B_x = B_{\rm in} \cos \chi~,~~~~B_y = B_{\rm in} \sin \chi~,
\nonumber
\end{eqnarray}
\begin{eqnarray}
v_x = -v_{\rm in} \cos \chi~,~~~~v_y = - v_{\rm in} \sin \chi~,
\nonumber
\end{eqnarray}
\begin{eqnarray}
\rho = \rho_{\rm in}~,~~~~p = p_{\rm in}~.
\nonumber
\end{eqnarray}
We consider that the pressure in the incoming flow is very small so
that it does not influence the  MHD-shock wave and cooling zone.
Equivalently, the sonic Mach number is much larger than unity. As
discussed earlier,  we choose a coordinate system in which the flow
velocity and magnetic field are both in the $(x,y)$ plane.

    In contrast with the non-magnetic case
 (TD93), the density $\rho$
does not increase unrestrictedly at the
right-hand boundary of the radiative zone and correspondingly the
velocity $v_x$ does not approach zero.
   Here, the temperature $T =
p/({\cal R} \rho)$ may approach zero not because $\rho \to \inf$ but
because $p \to 0$.
  At the same time the total, gas plus magnetic field pressure
remains constant.
  In the cooling zone the accreting
matter does not reach zero flow speed.
   Thus, the radiative shock provides
only part of the deceleration and absorption of
matter by the magnetized star.

     Our treatment of the
stability of the radiative MHD
shock wave is different in essential
respects from that of TD93 and RS05.
   In these
papers the magnetic field is transverse to the
flow and parallel to the star's surface, ${\bf B}=B_y\hat{\bf y}$.
   We neglect the difference in cooling laws and  adiabatic index
$\gamma$ (RS05) which are not significant.
  To approach this case in our model,
we need to take $\chi \to \pi/2$, and
parameters in the incoming flow $v_x = -v_{\rm in} \cos \chi,~ B_y =
B_{\rm in} \sin \chi \to B_{\rm in}$.
  The Alfv\'enic Mach number  is
\begin{eqnarray}
M_A = \frac{v_{\rm in} \cos \chi}{B_{\rm in}/ \sqrt{4\pi
\rho_{\rm in}}} = \frac{\cos \chi}{\sigma}~.
\end{eqnarray}
To have $M_A$ a fixed value as $\chi \to \pi/2$, we let
$\sigma \equiv \cos \chi /M_A$ and require
\begin{eqnarray}
B_x = B_{\rm in} \cos \chi \to 0~,\quad
v_x \to - M_A \frac{B_{\rm in}}{\sqrt{4\pi \rho_{\rm in}}}~,
\nonumber
\end{eqnarray}
\begin{eqnarray}
v_y = - M_A \tan \chi \frac{B_{\rm in}}{\sqrt{4\pi \rho_{\rm in}}}
\to \infty~.
\nonumber
\end{eqnarray}
  A  tangential velocity $v_y$ can be included in
the calculations of TD93 by a Galilean transformation
to another reference frame.
  Thus, a radiative MHD shock wave with the
magnetic field parallel to the star's surface
and perpendicular to the flow corresponds to the
limit where $ \chi \to
\pi/2$ and  $ \sigma = \cos \chi/M_A \to 0$.

\section{Dimensionless Variables and Scalings}

  Consider firstly the case where the radiation function is a power law,
$\Lambda(T) = A ({\cal R}T)^{\alpha}$, where ${\cal R}$ is the gas
constant. The coefficient $A$ has dimension
$ {\rm cm}^{5-\alpha}{\rm g^{-1}}{\rm s}^{\alpha-3}$.
For estimates  we assume that the hydrogen
and helium are completely ionized and that the average mass per particle
is $0.6m_p$.
   In the  paper GS07 the  radiative energy
losses per unit of volume are given in the form of equation (1).
With these simplifications and renormalization,
\begin{eqnarray}
\Lambda_{\rm GS} = 1.37 \times 10^{32} ({\cal R} T/{\rm
cm^2~s^{-2}})^{-0.54} {\rm cm^5~g^{-1}~s^{-3}} ~. \nonumber
\end{eqnarray}
We rewrite our equations in dimensionless form choosing fiducial
dimensional values for the main variables and introducing
dimensionless variables $\tilde A=A/A_0$ for different variables A.
The fiducial values are taken to be:
   For the velocity, $v_0 = v_{\rm in}\cos
\chi$; for the density, $\rho_0 = \rho_{\rm in}$; for the distance,
$v_0^{3-2\alpha}/A \rho_0$; for time, $v_0^{2-2\alpha}/(A \rho_0)$; for the
pressure, $\rho_0 v_0^2$; for the magnetic field, $v_0 \sqrt{\rho_0}$;
and for the temperature $v_0^2/{\cal R}$.
   In dimensionless variables, the system
of equations has the same form. Subsequently we remove tilde signs
from dimensionless variables.
    The formula for
the radiative losses now has the form $\rho^2 T^{\alpha}$, while in
the incoming flow $v_{\rm in} \cos \chi = -1, \rho_{\rm in} = 1,
B_{\rm in} = \sigma \sqrt{4 \pi},~ p_{\rm in} \ll 1$, where $\sigma$
is the inverse of the Alfv\'en Mach number.

  It is clear that in power law case,
$\Lambda \sim T^\alpha$, the stability/instability of
the radiative shock in presence of a magnetic field does not
depend solely on $v_{\rm in}$ and  $B_{\rm in}$, but instead on their
combination in the form $ (B_{\rm in}/\sqrt{4 \pi \rho_{\rm
in}})/v_{\rm in} = \sigma = M_A^{-1}$.
  Furthermore, the stability
criterion does not depend on the coefficient $A$.
    However, $A$ determines
the spatial scale of the radiative
zone and the temporal scale for
oscillations if they are present.

We chose the spatial scale is $ 10^7 {v_7^{4.08}}/(\rho_{-11}) {\rm
cm}$ and the time scale is $ {v_7^{3.08}}/(\rho_{-11}) {\rm s}$,
where $v_7$ is the velocity in units of $10^7 {\rm cm~s^{-1}}$ and
$\rho_{-11}$ is the density in units of $10^{-11} {\rm g~cm^{-3}}$.
In reality, both estimates are about three orders of magnitude
larger than the observed values. Such scales follow from the
solution of the equation for the stationary shock wave, where the
pressure goes to zero for $x>>1$ (TD93). The same is true for the
scale of time.
  Thus, the height of the cooling zone is in fact $\sim 10^4$ cm,
and the periods of oscillation are several hundredths of a second.

We now consider the ``real" radiative function
(RTV) which is not a power law.
This is why strictly speaking the  stability condition
of radiative shock waves depends not only on $\sigma$ but
also on both $v_{\rm in}$ and $B_{\rm in}$.
  However, over a
sufficiently wide range of temperatures,
the radiative function can be roughly
approximated by a power law of the temperature.
   This suggests that the stability condition
will depend mainly on $\sigma$ with a weaker
dependence on $v_{\rm in}$ and $B_{\rm in}$.

\begin{table*}
\begin{center}
\begin{tabular}{|c|c|c|c|c|c|} \hline
$N $ &$ cooling  $&$ v_{in} \cos \chi $&$ \sin \chi $&$ B_{in} $&$ stab /  $\\
     &$ function $&$ {\rm cm~s^{-1}}     $&$           $&$ (\rm G)    $&$ unstab  $\\ \hline
$I $ &$   RTV    $&$ 1.3 \times 10^7  $&$ 0.154     $&$  40.8  $&$
unst    $\\ \hline $II$ &$   GS     $&$ 1.3 \times 10^7  $&$ 0.154
$&$  40.8  $&$ unst    $\\ \hline $III$&$   RTV    $&$   3 \times
10^7  $&$ 0.154     $&$  94.9  $&$ unst    $\\ \hline $IV $&$   RTV
$&$ 1.3 \times 10^7  $&$ 0.368     $&$  36.4  $&$ stab    $\\ \hline
\end{tabular}
\caption{The Table shows parameters used in our main runs.}
\end{center}
\end{table*}

\section{Stationary structure}

We are interested in the stability of the stationary flow consisting
of the MHD shock wave and the downstream radiative zone.
  The stationary flow is described by
\begin{eqnarray}
\rho v_x = -j = - \rho_{\rm in} v_{\rm in} \cos \chi~,
\nonumber
\end{eqnarray}
\begin{eqnarray}
p + \rho v_x^2 +
\frac{B_y^2}{8\pi} = q_x = p_{\rm in} + \rho_{\rm in} v_{\rm in}^2
\cos^2 \chi + \frac{B_{\rm in}^2}{8\pi} \sin^2 \chi~,
\nonumber
\end{eqnarray}
\begin{eqnarray}
\rho v_x v_y - \frac{B_x B_y}{4\pi} = q_y = \left( \rho_{\rm in}
v_{\rm in}^2 - \frac{B_{\rm in}^2}{4\pi} \right) \sin \chi \cos
\chi~, \label{3}
\nonumber
\end{eqnarray}
\begin{eqnarray}
v_x B_y - v_y B_x = 0 ~,
\nonumber
\end{eqnarray}
\begin{equation}
j \frac{d}{dx}
\left( \frac{v^2}{2} + \frac{\gamma p/\rho}{\gamma-1} \right)
= \rho^2 \Lambda~.
\end{equation}
In the last equation there is no term describing the energy flux of
the electromagnetic field, because the Poynting flux in our
coordinate system is zero (${\bf E}=-{\bf v}\times {\bf B}/c =0$).

The stationary structure of the shock wave and
radiative zone is described
by the following equation for specific volume $V$,
\begin{eqnarray}
 \frac{dV}{dx} = \frac{\Lambda(T)}{j V^2 F(V)}~,
\end{eqnarray}
where
\begin{eqnarray}
 F(V) = j^2 V \left[ 1 + \frac{q_y^2 B_{x}^2/4\pi}
{(B_{x}^2/4 \pi - j^2V)^3} \right] +
\end{eqnarray}
\begin{eqnarray}
\quad \quad \frac{\gamma}{\gamma+1} \left[ q_x - 2j^2V - \frac{q_y^2
B_x^2/8\pi}{(B_x^2/4 \pi - j^2V)^3} \left( j^2V +
\frac{B_x^2}{4\pi}\right) \right]~.
\nonumber
\end{eqnarray}
The other variables are determined from equations (\ref{3}).
In particular,
\begin{eqnarray}
v_x = -jV~,~~~~B_y = -\frac{B_x q_y}{B_x^2/4\pi - j^2V}~,
\nonumber
\end{eqnarray}
\begin{eqnarray}
v_y = \frac{B_y}{B_x}v_x = \frac{q_y jV}{B_x^2/4\pi - j^2V}~,
\nonumber
\end{eqnarray}
\begin{equation}
p = q_x- j^2V - \frac{q_y^2 B_x^2/8\pi}{(B_x^2/4\pi -
j^2V)^2}~,~~~~ T = \frac{pV}{{\cal R}}~.
\label{4}
\end{equation}

As mentioned we investigate stability of the structure
``fast MHD shock wave"  plus ``cooling zone".
     Note that as the angle $\chi$
between the flow velocity and the normal to the
shock is decreased,
the stationary MHD structure does not convert to the
hydrodynamic one.
   In the limit $\chi \to 0$,
 $ q_x \approx {B_x^2}/(4 \pi \sigma^2)$.
   The pressure in the radiative zone is determined from the
relation
\begin{equation}
p \approx \frac{B_x^2}{4 \pi \sigma^2} - j^2 V - \frac{q_y^2
B_x^2/8\pi}{(B_x^2/4 \pi - j^2 V)^2}~,
\label{5}
\end{equation}
where $q_y \propto \sin \chi$  is small.
   For the considered conditions
where $\sigma < 1$, the denominator of the
fraction of the last term  goes to zero before the
$ {B_x^2}/({4 \pi \sigma^2}) - j^2V$ goes to zero.
   Thus the
pressure goes to zero for $B_x^2/4\pi - j^2V \approx 0$.
   That is, the
velocity of the flow at the exit from the radiative zone
approaches the Alfv\'en velocity and is $v_x \approx -B_x^2/(4\pi j)$.
   Taking into account equation (\ref{4}), we obtain from
equation (\ref{5}),
\begin{eqnarray}
0 \approx \frac{B_x^2}{4 \pi \sigma^2} - \frac{B_x^2}{4 \pi} -
\frac{B_x^2}{8 \pi} \left( \frac{v_y}{B_x^2/4\pi j} \right)^2~.
\end{eqnarray}
>From this relation we obtain
\begin{eqnarray}
v_y \approx - B_x \sqrt{\frac{1-\sigma^2}{2 \pi \rho_{\rm
in}}}~,~~~~ B_y \approx \frac{B_x}{\sigma} \sqrt{2 (1-\sigma^2)}~.
\end{eqnarray}
Thus, as $\chi \to 0$ the  components of the velocity
and magnetic field parallel to the shock
approach finite values.
     In contrast, in the gas dynamic case these
components approach zero.
 The observed behavior
is similar to that known to occur
in MHD switch-on shock waves where
finite tangential velocity and magnetic
field components are generated, and
where the velocity of the flow behind the front is Alfv\'enic
(Smith, 1993).
   Note that the parallel MHD shock wave becomes
non-evolutionary, and it is replaced by a switch-on
shock wave for $
[ {v_x}/(B_x/\sqrt{4 \pi \rho})]_{\rm in} < 2$ for
$\gamma = 5/3$ and  $ p_{\rm in} = 0$.

\begin{figure*}[t]
\epsscale{1.} \plotone{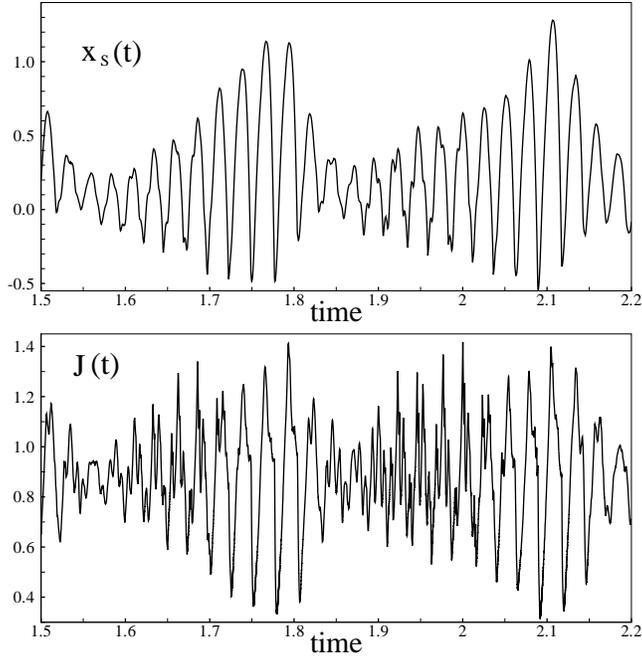} \caption{Time dependence of the shock
front coordinate $x_s$ (top panel) and luminosity $J$ (bottom panel)
for case I (``real" cooling function, $v_{\rm in} \cos \chi = 1.3
\times 10^7 {\rm cm~s^{-1}})$. The shock front coordinate is in
units $\Delta$, luminosity in units of $J_{\rm in}$, and time is in
seconds.} \label{Figure 4}
\end{figure*}
\begin{figure*}[b]
\epsscale{1.} \plotone{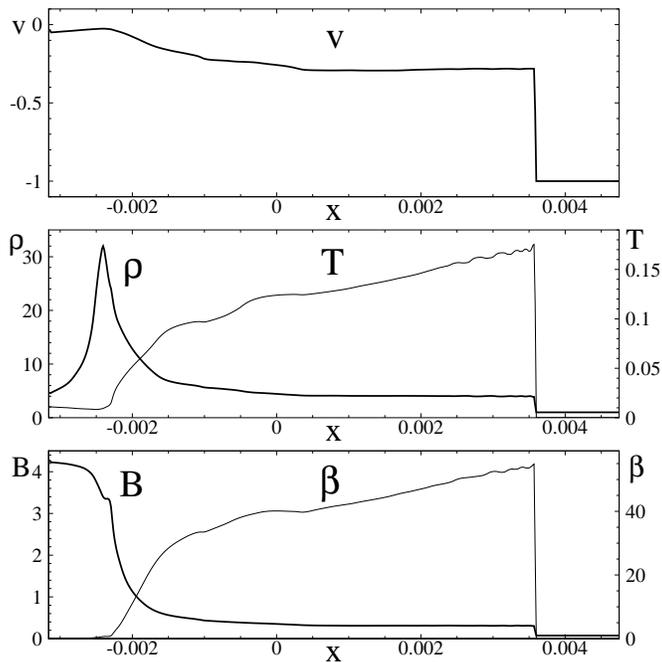} \caption{Distribution of different
parameters along the $x-$axis: velocity component along the field
lines (top panel), density and temperature (middle panel), magnetic
field and plasma parameter $\beta$ (low panel). All variables are
shown in dimensionless form. In the incoming flow $v_x=-1$,
$\rho=1$, $T=0$ (almost zero), $B_x=0.48$, $B_y=0.076$. Matter
inflows from the right boundary, a star is at the left.}
\label{Figure 5}
\end{figure*}
\begin{figure*}[t]
\epsscale{1.} \plotone{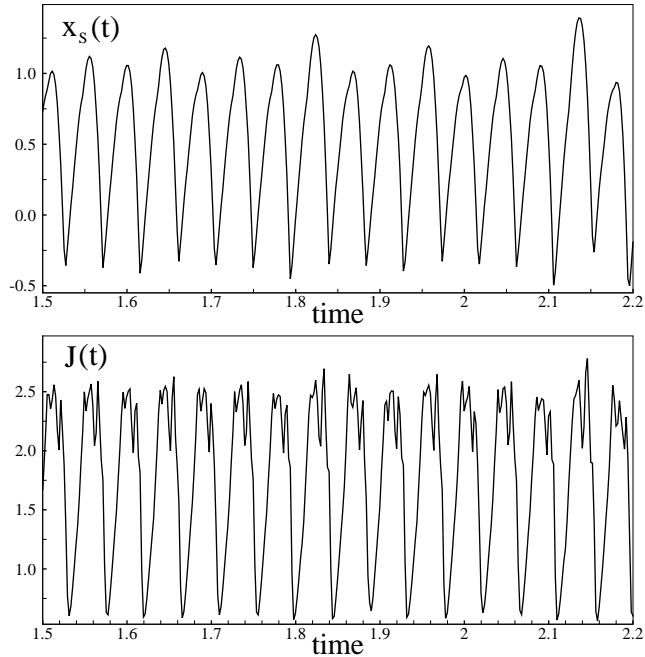} \caption{Time dependence of shock
front coordinate $x_s$ (top panel) and luminosity $J$ (bottom panel)
for case II (power-law cooling function, $v_{\rm in} \cos \chi = 1.3
\times 10^7 {\rm cm~s^{-1}})$. The coordinate of the shock front
$x_s$  is in units $\Delta$, the luminosity is in units  of $J_{\rm
in}$, and the time is in seconds.} \label{Figure 6}
\end{figure*}
\begin{figure*}[b]
\epsscale{1.} \plotone{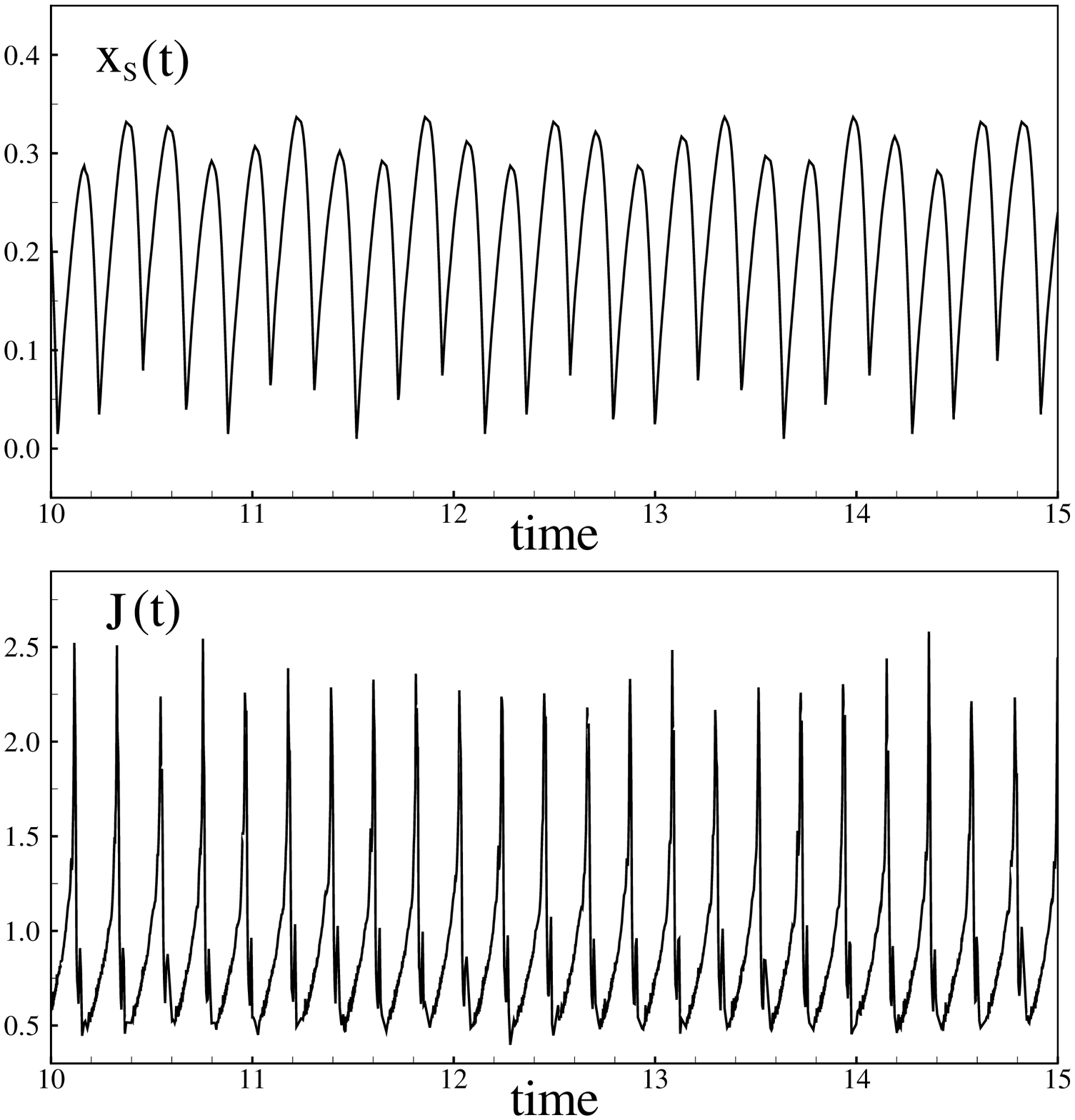} \caption{Time dependence of the shock
front coordinate $x_s$ (top panel) and luminosity $J$ (bottom panel)
for case III (``real" cooling function, $v_{\rm in} \cos \chi = 3
\times 10^7 {\rm cm~s^{-1}}$. The shock front coordinate is in units
$\Delta$, the luminosity is in units of $J_{\rm in}$,  and the time
is in seconds.} \label{Figure 7}
\end{figure*}
\begin{figure*}[t]
\epsscale{1.} \plotone{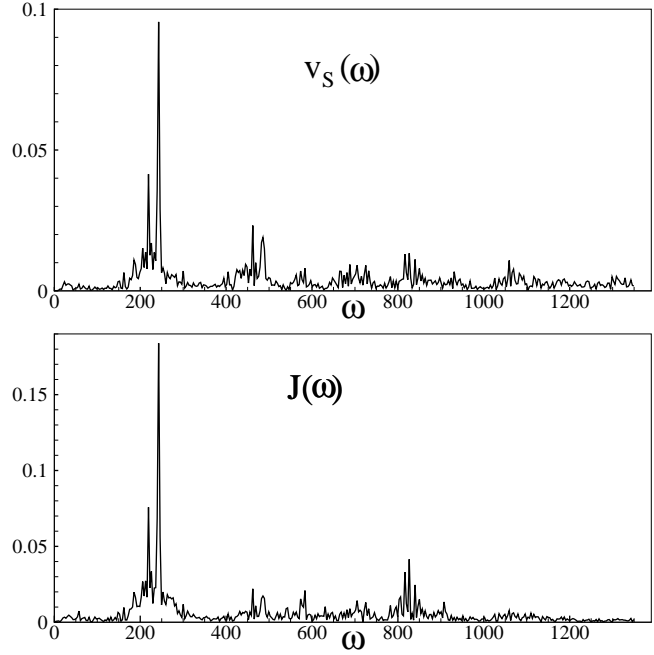} \caption{Fourier amplitudes of shock
front velocity $v_s$ (top panel) and luminosity $J$ (bottom panel)
for case I. Velocity is in $v_{\rm in} \cos \chi$, luminosity is in
$J_{\rm in}$, frequency is in ${\rm s}^{-1}$.} \label{Figure 8}
\end{figure*}
\begin{figure*}[b]
\epsscale{1.} \plotone{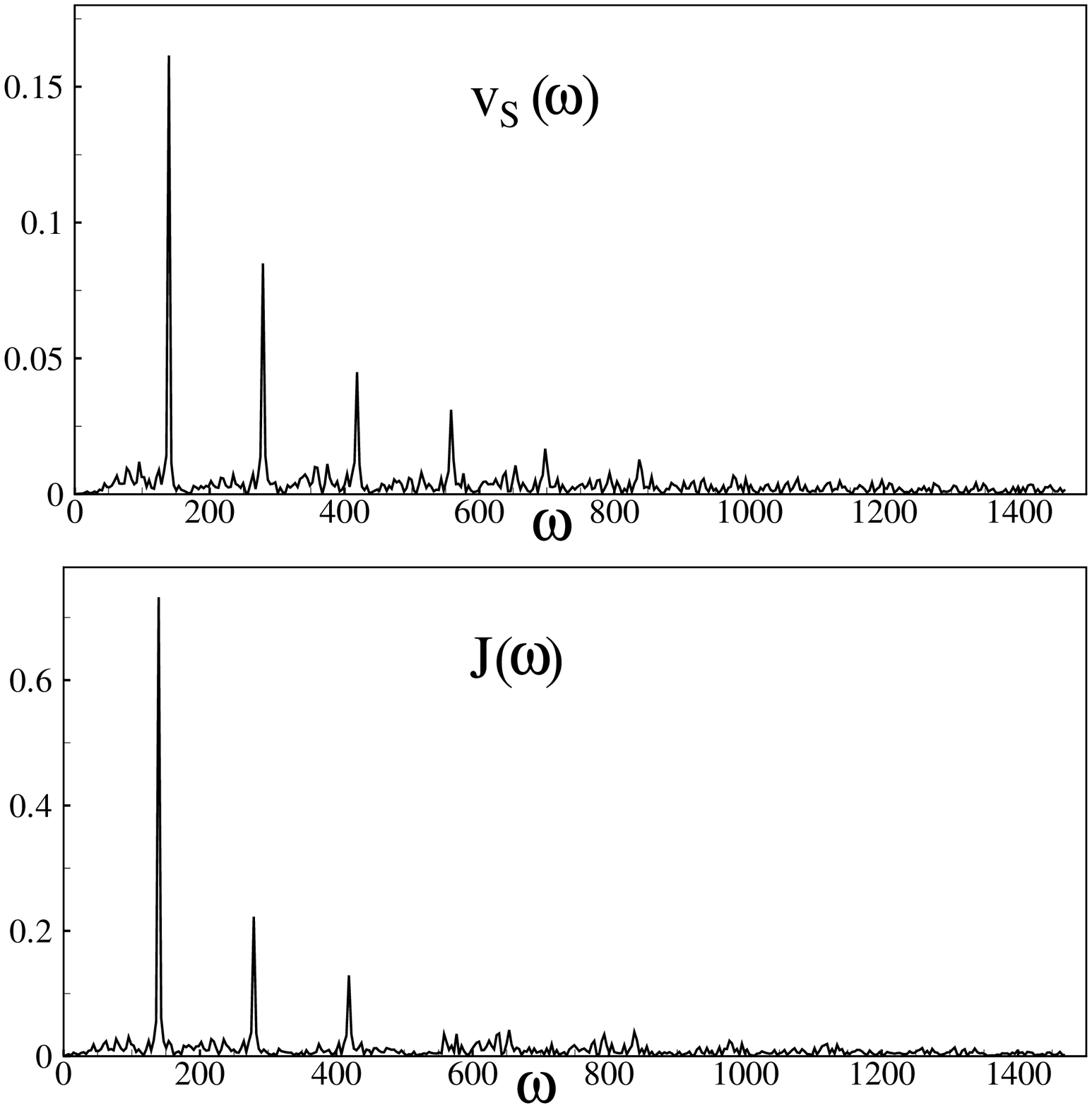} \caption{Fourier amplitudes of shock
front velocity $v_s$ (top panel) and luminosity $J$ for case II
(cooling function $\Lambda_{\rm GS}$, $v_{\rm in} \cos \chi = 1.3
\times 10^7 {\rm cm s^{-1}})$). The velocity is in units of $v_{\rm
in} \cos \chi$, the luminosity is in units of $J_{\rm in}$, and
frequency is in ${\rm s}^{-1}$.} \label{Figure 9}
\end{figure*}

\begin{figure*}[t]
\epsscale{1.} \plotone{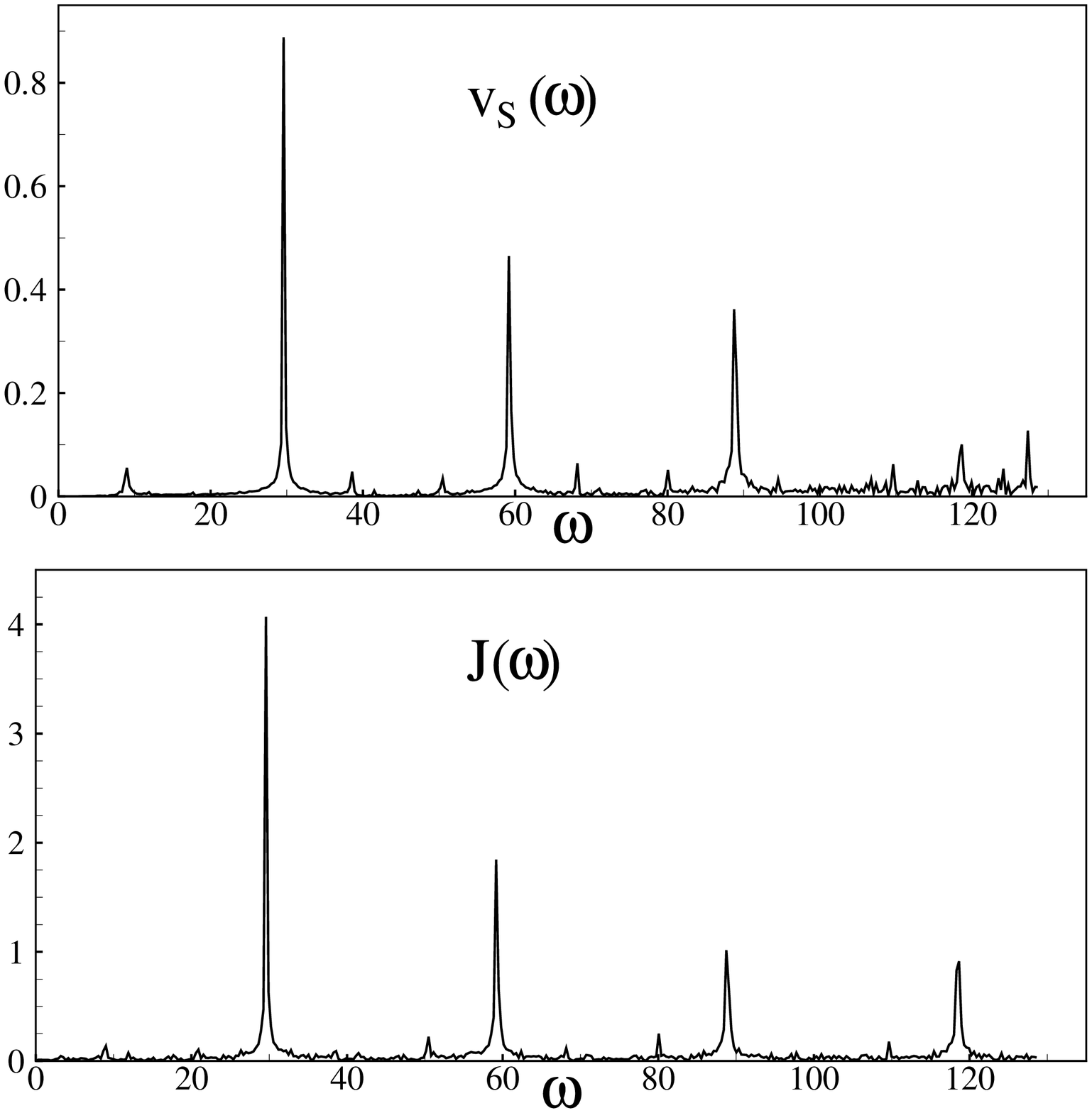} \caption{Fourier amplitudes of shock
front velocity $v_s$ and luminosity $J$ for case III (cooling
function $\Lambda_{\rm RTV}$ $v_{\rm in} \cos \chi = 3 \times 10^7
{\rm cm s^{-1}})$). The velocity is in units of
 $v_{\rm in} \cos \chi$, the  luminosity is in units of
$J_{\rm in}$, and the frequency is in ${\rm s}^{-1}$.} \label{Figure
10}
\end{figure*}

\begin{figure*}[b]
\epsscale{1.} \plotone{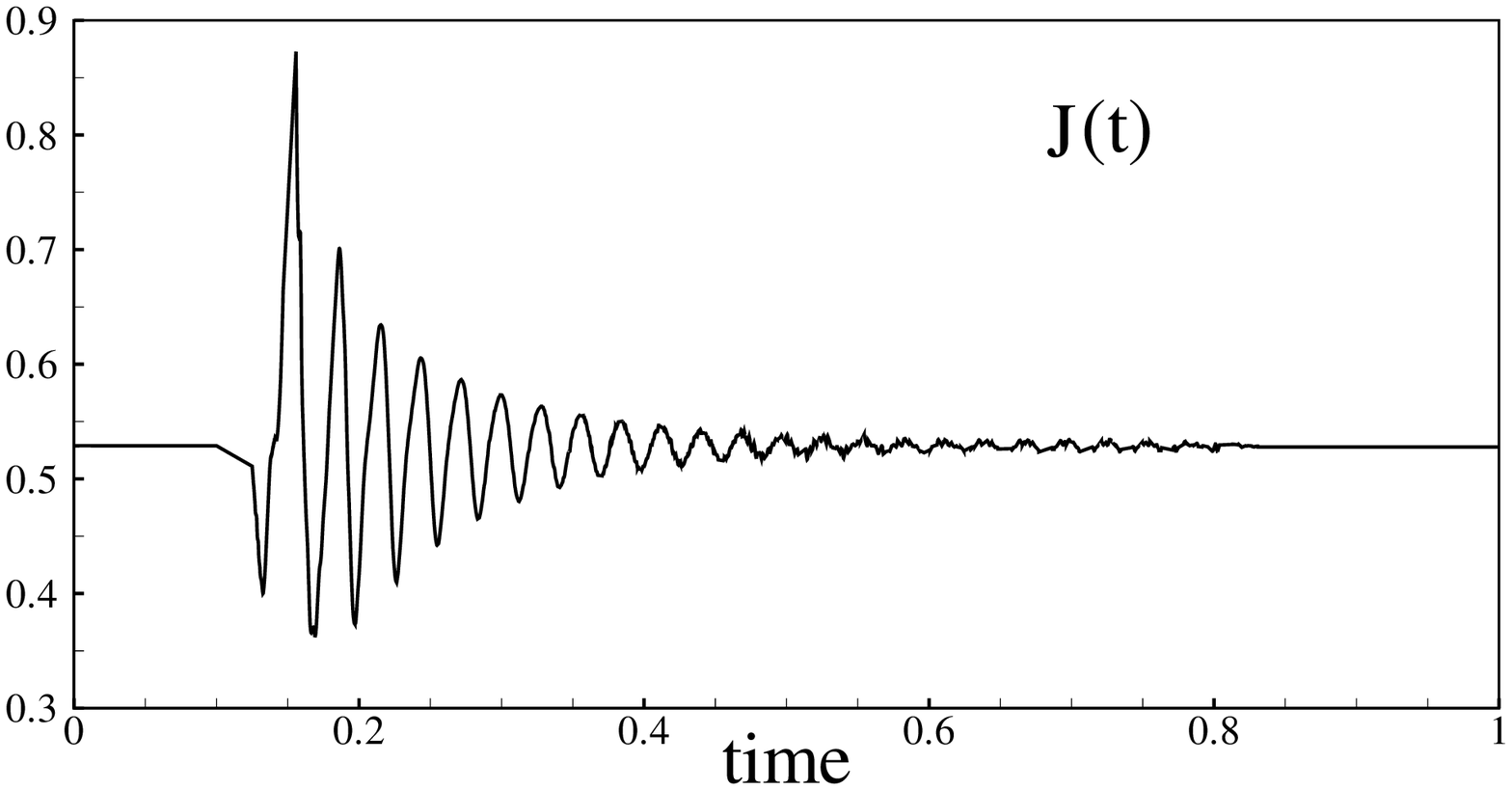} \caption{Time dependence of
luminosity $J$ for case IV (cooling function $\Lambda_{\rm RTV},
v_{\rm in} \cos \chi = 1.3 \times 10^7 {\rm cm~s^{-1}})$. The
luminosity is in units  of $J_{\rm in}$, and the time is in
seconds.} \label{Figure 11}
\end{figure*}
\section{Method}

   We study the stability/instability of radiative
shock waves in the presence of the
magnetic field by integrating the
time-dependent one-dimensional MHD equations in a region
containing the shock and the radiative cooling zone.
  We used an Eulerian
variables.
    Consequently, the simulation region is chosen large
enough to contain
 both the shock wave and the remote part of the
radiative zone where energy losses are negligibly small.
  The location of the right-hand
boundary of the simulation region is
chosen so that
the shock wave did not leave the region during
oscillations.
  For the calculations of the MHD flows we used
a   high resolution Godunov type numerical scheme (see, e.g.,
Kulikovskii, Pogorelov \& Semenov 2000).

For initial conditions we take the stationary flow with the shock
wave and the radiation zone. At the top of the simulation region
(see Figure 1) we determined either parameters of unperturbed flow
(density, velocity, etc. ) if the incoming flow is super-Alfv\'enic,
or we had ``free" boundary conditions, if incoming flow is
sub-Alfv\'enic.

At the bottom of the simulation region (closer to stellar surface),
we have a ``layer" with no radiative losses. At this boundary we
fixed the longitudinal velocity at a small value corresponding to
the stationary solution. All other variables at the boundary had the
same value as in the previous cell. Simulations have shown that
results of modeling are not sensitive to the boundary conditions on
the transverse components of velocity and magnetic field. At the
boundary cell the density varied around some average value and the
mass did not accumulate there. For example in case I the sound speed
in this cell has been $(1-3)\times 10^6 {\rm cm~s^{-1}}$, the size
of the grid $=81 {\rm cm}$, the sound-crossing time $<10^{-4} {\rm
s}$. We observed from simulations that variation of this boundary
did not influence much to the oscillations.

The spatial resolution has been chosen so that the radiation zone is
covered by 200 cells. The size of the simulation region has been
chosen such that during the oscillations the shock wave stayed
inside the simulation region. Depending on the amplitude of
oscillations the simulation region incorporated from $500$ to
$1,500$ cells. The time-step has been chosen automatically such that
the Courant number is $0.5$.

\section{Results}

     We investigate the stability/instability
of the radiative shock wave as a
function of two main
dimensionless parameters: the inverse
Alfv\'en Mach number
of the unperturbed upstream flow $\sigma =
M_A^{-1}$ and the inclination angle of the flow, $\chi$,
relative to the normal to the star's surface
  The limit $\sigma = 0$ corresponds to zero magnetic field.

We performed a series of calculations
in which we varied these two
parameters.
   Figure 3 summarizes the results.
In the plane $(\sigma, \sin \chi)$ markers show the
parameters where calculations were done.
    The ``squares"
show the parameters for which the radiative shock wave is stable
and the ``triangles" show cases where it is unstable.
   The solid straight
line with a dashed continuation is the stability/instability
boundary.
    The condition for stability  can be expressed as
\begin{equation}
3.7 \sigma + 1.4 \sin
\chi \ge 1~,
\end{equation}
for conditions where the magnetic
field is not very weak, $\sigma \ge 0.02$.

   For weak magnetic fields ($\sigma \le 0.02$),
the boundary of stability is located close to the vertical axis in
Figure 3.
   This part of the boundary is shown as a
dashed line.
   Thus, in the absence of a magnetic field, the radiative shock
wave (for the same parameters of the incoming flow) is unstable.
However, even a small magnetic field, in particular, inclined one,
stabilizes the radiative shock wave.
   For $\sigma \ge 0.02~~(M_A \le 50)$,
we find stability for angles $\chi > 45^\circ$.
  If the magnetic field is sufficiently
strong $\sigma \ge 0.27~~(M_A \le 3.7)$,
then the radiative shock wave is stable for all angles $\chi$.

We emphasize again that in the limit of small
inclination angles, $\chi \to 0$, our model
does not correspond to a
gas-dynamical flow with unperturbed magnetic field.

   To understand the temporal
characteristics of unstable radiative shock
waves, we performed calculations using
the ``real'' cooling function $\Lambda_{RTV}$
and for some cases with the
function $\Lambda_{\rm GS}(T)$.
   In all calculations we followed the
location of the shock wave $x_s(t)$ and the total intensity of radiation
from the radiative zone per unit area $J(t) = \int dx \rho^2 \Lambda$.

Figure 4 shows results of calculation of evolution of the radiative
shock wave with radiation function $\Lambda_{\rm RTV}$ for the
following parameters of the incoming flow: $v_{\rm in} = 1.3 \times
10^7 {\rm cm~s^{-1}},~\rho_{\rm in} = 10^{-11} {\rm g
cm^{-3}},~B_{\rm in} = 40.8 {\rm G},~\sin \chi = 0.154$.
   Also,  $\sigma = 0.13$.
The width of the cooling zone in the stationary regime is $\Delta
\approx 1.64 \times 10^4 {\rm cm}$, and the energy-density
(excluding the much smaller thermal energy) is $J_{\rm in} = (1/2)
\rho_{\rm in}~v_{\rm in}^3 \cos \chi = 1.07 \times 10^{10} {\rm
erg~cm^{-2}~s^{-1}})$.
  We  point out once again that the incoming flow
velocity is  along the magnetic field, so that there is no
Poynting flux.

   For the mentioned parameters, the stationary shock is unstable
and the shock and radiative zone oscillate.
     The position of the
front of the shock wave oscillates with a period $\approx 0.025$ s.
The amplitude of oscillations of the front is modulated and varies
with period $\approx 0.3 $ s.
    The top panel of Figure 4  shows the
position of the front as a function of time at the time-interval
approximately equal to twice period of modulation.
   The bottom panel of Figure 4 shows
the temporal variation of the radiation
intensity from the radiation zone
for the same time-interval as the top panel.
  The position of the
shock  front is normalized to the width of the stationary cooling zone
$\Delta$.
   The intensity of radiation is normalized to the energy-density in the
incoming flow $J_{\rm in}$.  The time is in seconds.

Figure 5 shows spatial distribution of the longitudinal
($x-$direction) velocity (top panel), density and temperature
(middle panel), longitudinal magnetic field and plasma parameter
$\beta=8\pi p/({B_x^2+B_y^2})$ (bottom panel) at a  time
corresponding to the maximum distance of the shock from the surface
of the star.
  In stationary regime the shock wave is located at $x=0$,
while the radiative zone is below this. See animations at
http://www.astro.cornell.edu/us-rus/shock.htm

Figure 6 shows results of calculation of the evolution of the
radiative shock wave for the same parameters in the incoming flow
but with the radiation function $\Lambda_{\rm GS}$ (case II).
   The width
of the cooling zone in the stationary regime is $\Delta \approx 2.65
\times 10^4 {\rm cm}$.
  This figure shows position of the front of the shock wave $x_s(t)$ (top
panel) and intensity of radiation $J(t)$ (bottom panel).
   One can see
that qualitatively the  oscillations are similar.
   However,
the main frequency of oscillations is different: $\omega_{\rm RTV}
\approx 240 {\rm s^{-1}},$ whereas  $ \omega_{\rm GS} \approx 140 {\rm
s^{-1}}$.
    With the radiation function $\Lambda_{\rm
GS}$, the modulation of the oscillation amplitude of the position of
the front is small.

Figure 7 shows results of calculation of evolution of the radiative
shock wave with radiation function $\Lambda_{\rm RTV}$ for the
following parameters of the incoming flow (case III): $v_{\rm in} =
3 \times 10^7 {\rm cm~s^{-1}},~\rho_{\rm in} = 10^{-11} {\rm g
cm^{-3}},~B_{\rm in} = 94.9 {\rm G},~\sin \chi = 0.154$. Also,
$\sigma = 0.13$. The width of the cooling zone in the stationary
regime is $\Delta \approx 8.86\times10^5 {\rm cm}$, and the
energy-density (excluding the small thermal energy) is $1.35 \times
10^{11} {\rm erg~cm^{-2}~s^{-1})}$.
   As in case I for these parameters the
stationary shock is unstable and the shock oscillates.
    The position of the front of the shock wave
oscillates with period $\approx 0.21 {\rm s}$.
   The top panel shows the
position of the front as a function of time,
and the bottom panel shows the
variation of the radiation intensity
for the same time interval as the top panel.

In case III we increased both the velocity and magnetic field with
the aim of checking our hypothesis, that the qualitative solution of
the problem (while using the ``real" radiation function
$\Lambda_{\rm RTV}$) is determined by the dimensionless parameters
of the problem, $\sin \chi$, and the magnetization, $\sigma$, and
not by dimensional values of velocity of the incoming flow and the
magnetic field.

Figure 8 shows results of the Fourier-analysis of the speed of the
shock front and total luminosity per unit of area $v_s(\omega)$ and
$J(\omega)$. One can see that when modelling with radiation function
$\Lambda_{\rm RTV}$ (case I) then two nearby maxima of the Fourier
amplitudes are observed at the frequencies $\omega_1 = 240 {\rm
s^{-1}},~ \omega_2 = 260 {\rm s^{-1}}$.
  We suggest that  the combination of these two frequencies
gives the amplitude modulation evident in Figure 4.

Figure 9 shows the Fourier amplitudes for case II where we use
$\Lambda_{\rm GS}$.
   The Fourier-amplitudes for both, $v_s$, and $J$
have sharp maxima at frequencies divisible by the main frequency.

Figure 10 shows the Fourier amplitudes for case III where we use
$\Lambda_{\rm RTV}$. The Fourier-amplitudes for both, $v_s$, and $J$
have sharp maxima at frequencies divisible by the main frequency.
The highest peak in Figures 8-10 corresponds to the main oscillation
frequency of the shock.
    The other peaks correspond to higher harmonics due
the oscillations being anharmonic.

We also investigated the stable regime of the radiative shock wave
(case IV). To study  the damping of the oscillations we introduced a
small ($10\%$) perturbation of the velocity in the upstream flow
during a limited time. We used the ``real" radiative function, and
parameters as shown in the Table for case IV. One can see that when
perturbations reached the front of the shock wave, the shock wave
began to oscillate. However, these oscillations damped during
several periods and stationary flow was
 re-established.  Figure 11 shows an example of such
damping.

\section{Conclusions}

This work has studied the
stability/instability of the radiative
MHD shock waves in the funnel
streams of classical T~Tauri stars.
Matter flowing to the surface
of the star along the magnetic field is decelerated in the
radiative shock wave.
The shock may be stable or unstable.
In the case of instability the shock position and other
variables oscillate and this can give observable
short time-scale variability in the emitted radiation.
  A significant new aspect of the present work is
that the magnetic field and the flow velocity parallel to it can
have an arbitrary angle with respect to the normal to the star's
surface.

     The shock wave has been modeled by solving the
time-dependent MHD equations in one dimension (perpendicular
to the star's surface) taking into account the radiative losses.
   For the radiative losses we used either the ``real"
radiative function, approximated by segments of power laws
(RTV) or by the power law function proposed by GS07.

Results of modelling of the radiative shock waves show that there is
a simple criterion of the shock stability: $3.7 \sigma + 1.4 \sin
\chi > 1$.  This is for the case where the inflow  to the shock is
$v_{\rm in} \cos \chi = 1.3 \times 10^7 {\rm cm~s^{-1}}$.
    We believe that  this criterion will not
change significantly at larger inflow velocities.

Comparison of the simulation results with the ``real" (RTV) and power-law
(GS) radiative functions  shows that the qualitative results are
similar.  However, the periods of oscillations are significantly
different.

The periods of oscillations are of the order of hundredths of a
second for  $v_{\rm in} \cos \chi = (1.3-3.0) \times 10^7 {\rm
cm~s^{-1}}$.
  This period is expected to increase with  $v_{\rm
in}$.  We estimate that $ P \approx 6 \times 10^{-3} [(v_{\rm in}
\cos \chi)/(10^7 {\rm cm~s^{-1}})]^3 {\rm s}$. The period of the
oscillations varies, $P=0.02-0.2$, depending on parameters.

Global three-dimensional simulations of magnetospheric accretion
through the funnel streams have shown that hot spots on the surface
of the star are {\it not homogeneous}: most of the kinetic energy
flows in the central regions of the funnel stream so that the
central regions of the spots are expected to be hotter (and also
denser) compared to peripheral regions (Romanova et al. 2004;
Kulkarni \& Romanova 2005). Romanova et al. (2004) have shown that
the spots may have very small filling factor at highest density and
temperature (less than $1\%$) and much larger filling factor at
smaller densities and temperatures (see Fig. 3 of Romanova et al.
2004). This fact has been recently confirmed observationally by
G\"unther et al. (2007) who have shown that the filling factor in
X-ray is smaller compared with UV and optical bands which confirmed
the  inhomogeneity of the hot spots. Future research should be done
for analysis of the stability of the global shock wave which would
cover a significant part of the hot spot (not a small part as
usually considered including this paper).

If the magnetic field near the star has a  complex geometry (e.g.
Valenti \& Johns-Krull 2004; Gregory et al. 2006; Donati et al.
2007;  Long, Romanova \& Lovelace 2007, 2008), then it is likely
that some field lines are inclined to the surface of the star as
considered in this paper.    The present analysis is thus applicable
to the stability/instability of the shocks.  If the complex field
has significant transverse component then it may suppress
oscillations.

Kravtsova \& Lamzin (private communication) searched for
oscillations of the shock in RW Aur using Crimean Observatory
facilities. They did not find oscillations, though their
time-resolution was low ($\Delta P=0.5 - 1 {\rm s}$. They concluded
that in this star there are no oscillations with periods $P > 2 {\rm
s}$ with amplitudes $> 5\%$ above the noise level. Higher
time-resolution observations in a larger sample of CTTSs are needed
to obtain a conclusive answer.
   It would be useful to
have high ({\rm ms}) time-resolution observations in the UV and
 X-ray bands in stars with high veiling, such as RW Aur and
others, because these wavebands would correspond to oscillations of
the central part of the hot spots (Romanova et al. 2004; G\"unter et
al. 2007) which may go into global oscillation mode with higher
probability compared to peripheral parts observed in the optical
band.
   A search
in the optical band may bring interesting results as well because we
do not know the details  of the interaction of the funnel stream
with a star on a global scale of the size of hot spot. To understand
such physics, observations of variability  time-scales would be
informative and may help to shed a light to process of the
funnel-star interaction.

\section*{Acknowledgements}

Authors thank Dr. Lamzin and Dr. Beskin for helpful discussions and
an anonymous referee for multiple questions and comments which
improved the paper. This work has been partially supported by the
NSF grants AST-0507760 and AST-0607135, and NASA grants NNG05GG77G
and NAG5-13060. A.V.K. and G.V.U. were partially supported by RFBR
grant 06-02-16608.

 \end{document}